# ANISOTROPY AND CRYSTALLITE MISALIGNMENT IN TEXTURED SUPERCONDUCTORS

D. M. Gokhfeld[1,2], S. V. Semenov[1,2], M. I. Petrov[1], I. V. Nemtsev[1,2,3], and D. A. Balaev[1,2]

[1] Kirensky Institute of Physics, Federal Research Center KSC SB RAS, Krasnoyarsk, 660036 Russia

[2] Siberian Federal University, Krasnoyarsk, 660041 Russia

[3] Federal Research Center "Krasnoyarsk Science Center of the Siberian Branch of the Russian Academy of Sciences", Krasnoyarsk, 660036, Russia

A misalignment of anisotropic crystallites causes small values of anisotropy and decreases the critical current density of textured polycrystalline superconductors. To relate the crystallite misalignment and out-plane anisotropy, the magnetic properties of the textured Bi2223 polycrystalline superconductor were investigated. A distribution of orientation angles of crystallites was determined using different data: scanning electron microscopy images and hysteresis magnetization loops when an external magnetic field was applied at different angles with respect to the texturing plane of the sample. It was demonstrated that the standard deviation of the distribution and the magnetic disorder angle of crystallites in textured samples can be determined from the magnetization data in perpendicular directions. These data may be either the irreversible magnetization measured for two different orientations of the sample or the simultaneously measured magnetization projections parallel and perpendicular to the magnetic field.

1. Introduction

The outstanding values of the critical current density in the *ab* plane $j_{c,ab}$ and the high critical temperatures could bring $Bi_2Sr_2CaCu_2O_{8+x}$ (Bi2212) and $Bi_2Sr_2Ca_2Cu_3O_{10+x}$ (Bi2223) at the top of superconductor applications. However, the strong anisotropy coefficient $\gamma \gtrsim 100$ of these materials [1] hinders their rise. The overall critical current density $j_c$ of polycrystalline superconductors is very sensitive to arrangement of crystallites [2]. An alignment is crucially important for textured superconductors and tapes [3–8]. The crystallite misalignment strongly decreases $j_c$ as well as $\gamma$. Bi2212 and Bi2223 textured ceramics usually have low $\gamma$ and low $j_c$ values (e.g., $\gamma \approx 1.7$-$2.5$ and $j_c \approx 0.1$~$1$ kA/cm$^2$ at $T = 77.4$ K [9–13]) as compared with single crystals and wires. Optimization of the current carrying capacity in textured superconductors requires information about a distribution of crystallite orientation angles and reasons of the misalignment.

In anisotropic type-II superconductors, the vortex lattice and magnetization depend on an angle between the external magnetic field and the crystal principal axes [14–17]. The magnetization of textured polycrystalline sample is resulted from collective response of the crowd of crystallites. So the degree of texture in these samples is essential. The degree of texture is estimated from X-ray diffraction data (θ-2θ and ω-scans) [7, 18, 19]. In recent work [13], the integral magnetic method was suggested. The magnetic misalignment angle $θ^*$ was introduced to characterize texture in the Bi2223 superconductor. This angle $θ^*$ equals an averaged value of the modulus of crystallite deviation from the texturing plane. In presented work, new ways are paved to determine the distribution of the crystallite orientation angles and the corresponding magnetic misalignment angle $θ^*$ in the textured superconductors.

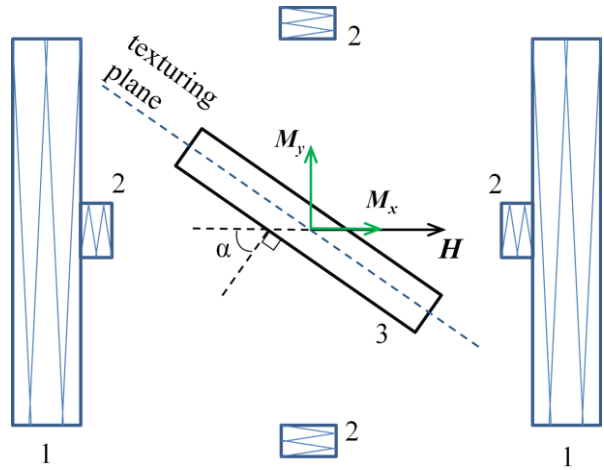

Fig. 1. Schematic for magnetization measurements. The numbers mark the magnet coils (1), the measuring coils (2), and the sample (3).

## 2. Experimental methods

The textured Bi2223+Ag ceramics was produced by two step route [9]. At the first stage the polycrystalline ceramics were synthesized from $Bi_2O_3$, $SrCO_3$, PbO, CuO, CaCO, and Ag powder by the solid-state synthesis. Porosity of the synthesized material is about 60% [20]. At the second stage the texture was created: the porous material was impregnated by ethanol, pressed up to 500 MPa, and annealed at 830 °C for 50 hours.

Scanning electron microscopy (SEM) was carried out using a Hitachi TM4000Plus microscope. Magnetization was measured using a Lakeshore VSM 8604 vibrating sample magnetometer. Measurements were carried out at 77.4 K for different orientations of the rotated sample relative to the external magnetic field ***H***. The sample had a parallelepiped form and sizes of 10×1×1 $mm^3$. The wider sides of the sample corresponded to the texturing plane. A schematic for magnetization measurements is shown on figure 1. Two magnetization projections were measured: ***$M_x$*** along the ***H*** direction (it is a general magnetization) and ***$M_y$*** perpendicular to ***H***.

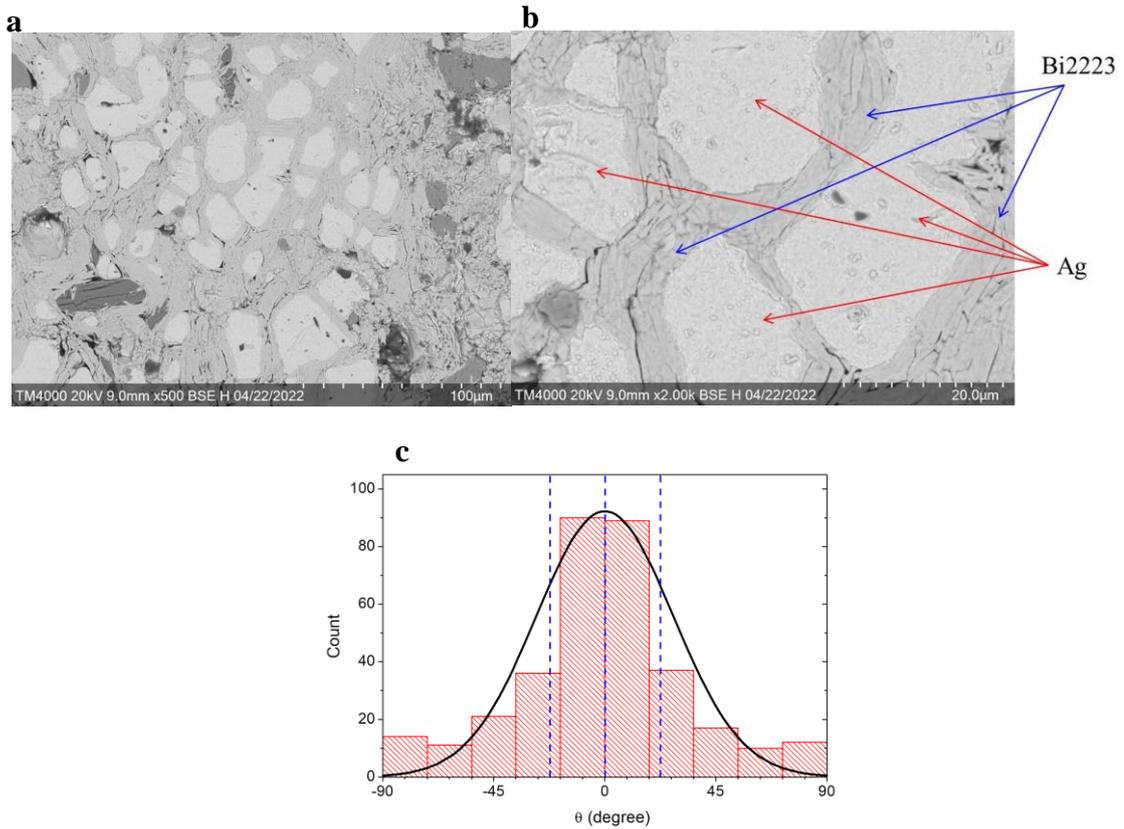

Fig. 2. Crystallites in the textured sample. a), b) SEM images of the sample side perpendicular to the texturing plane. c) Distribution of crystallite orientation angles. The bar chart is data obtained from SEM images. The normal distribution curve is computed (3) for σ = 28°. Vertical dash lines separate the quartiles.

### 3. Experimental results

SEM images of the sample side, which is perpendicular to the texturing plane, are shown on figures 2a,b. As seen, light particles of Ag are surrounded by darker crystallites of Bi2223. The diameter of the silver particles is about 20 μm. The Bi2223 crystallites have a flake-like form with the sizes about 10 μm × 10 μm × 1 μm. The most of crystallites are oriented along the texturing plane. However some of the crystallites are tilted because they go around the silver particles. We measured the values of the tilt θ of each crystallites on four different SEM images. The obtained distribution of crystallite orientation angles is presented on figure 2c. The distribution has a characteristic bell-like form.

The angle dependencies of the magnetization projections at $H$ = 5 kOe, which is much higher than the irreversibility field, are shown on figure 3a (α changes from –180° to +180° at this figure). The period of these dependencies is equal to 180°. It can be seen, the positions of $M_x$ = 0 do not correspond to the extremum positions of $M_y$.

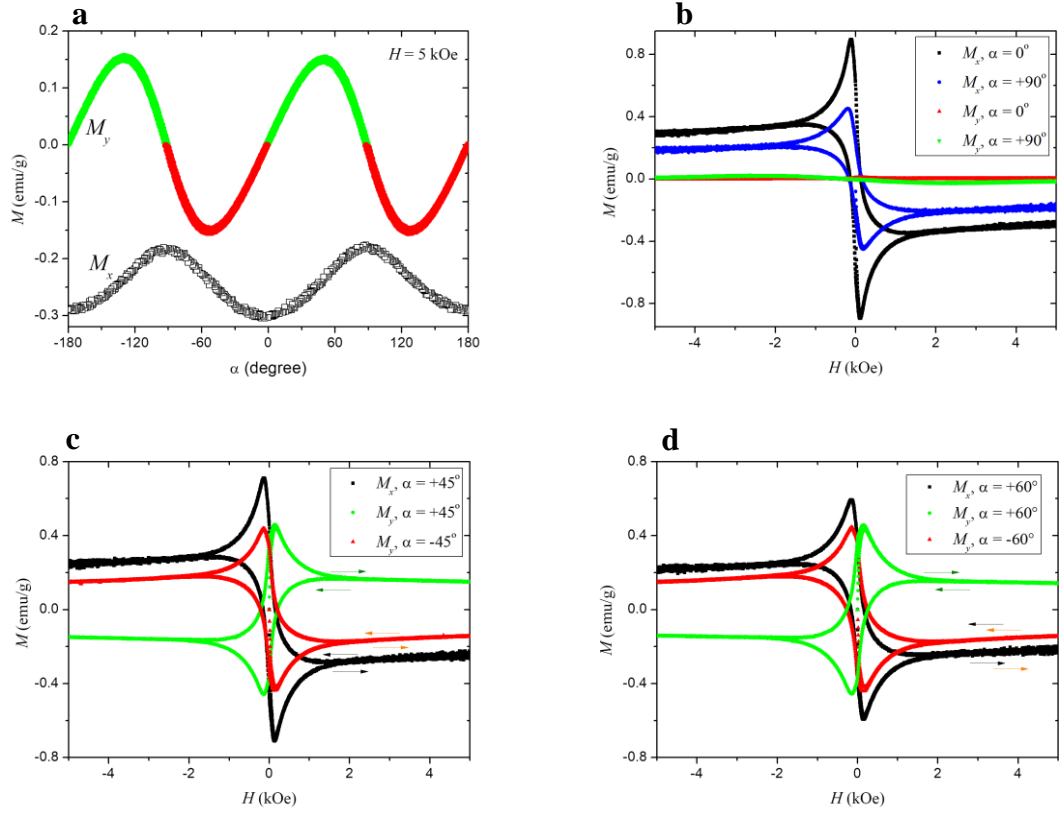

Fig. 3. Experimental magnetization isothermal curves at $T$ = 77.4 K. a) Angle dependencies of magnetization projections at $H$ = 5 kOe. Magnetization hysteresis loops for $M_x$ and $M_y$ projections b) α = 0° and +90°, c) α = -45° and +45°, d) α = -60° and +60°. The arrows mark directions of the change of $H$.

Figures 3b,c,d show magnetization hysteresis loops (the $M_x$ and $M_y$ projections) for some values of the angle α between the texturing plane and $H$. It is seen, the $M_y$ loops are inverted at the angle range 0° < α < +90°.

Further we focus on the dependence of the irreversible magnetization $M_{irr}$ on α. The irreversible magnetization is determined from the magnetization hysteresis loops as $M_{irr}(H)$ = $|M\downarrow(H) - M\uparrow(H)|/2$, where $M\uparrow(H)$ and $M\downarrow(H)$ are the magnetization curves at increasing and decreasing magnetic fields correspondingly. The $M_{irr}(H)$ dependencies for some values of α are plotted on figure 4a.

We designate $M_{irr}(H=0)$ at some α as $M_{irr,x}(α)$ and $M_{irr,y}(α)$ for the $M_x$ and $M_y$ projections correspondingly. The experimental loops for different α demonstrate that $M_{irr,x}$ decreases in about 2.5 times as α changes from 0° to +90° (or from 0° to -90°). The value of $M_{irr,y}$ is almost zero at α = 0 and α = ±90°. The maximal values of $M_{irr,y}(α)$ are achieved at α = -70° and α = +70°, for these angles $M_{irr,y} ≈ M_{irr,x} ≈ 0.6 M_{irr,x}(0°)$.

It is convenient to trace three normalized magnetizations $k_{xx0}(α) = M_{irr,x}(α)/M_{irr,x}(0)$, $k_{yx0}(α)$ = $M_{irr,y}(α)/M_{irr,x}(0)$, and $k_{yx}(α) = M_{irr,y}(α)/M_{irr,x}(α) = k_{yx0}(α)/k_{xx0}(α)$. These values obtained from the experimental magnetization hysteresis loops are shown on figures 4b,c,d. A model describing these dependencies is considered at next section.

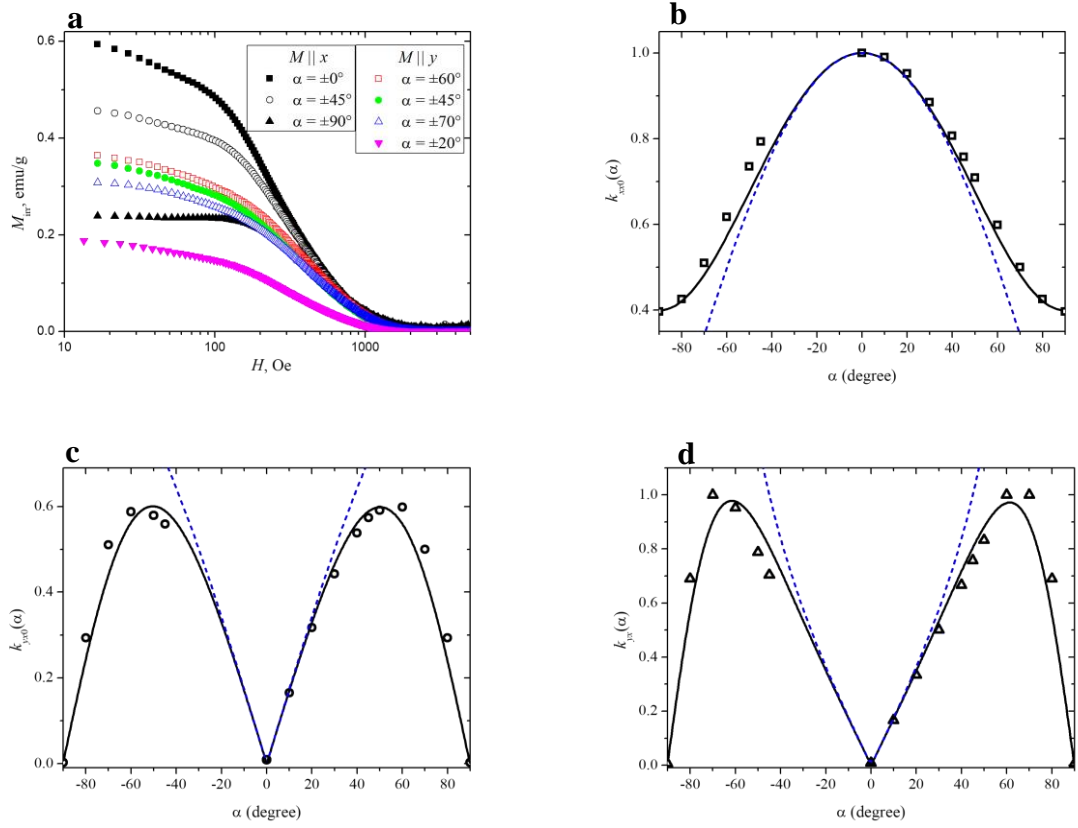

Fig. 4. The irreversible magnetization. a) The dependence of $M_{irr}$ on the magnetic field $H$ (The axis H is in a logarithmic scale). The normalized values of irreversible magnetization: b) $k_{xx0} = M_{irr,x}(\alpha)/M_{irr,x}(0)$, c) $k_{yx0} = M_{irr,y}(\alpha)/M_{irr,x}(0)$ and d) $k_{yx} = M_{irr,y}(\alpha)/M_{irr,x}(\alpha)$. Symbols are obtained from experimental magnetization hysteresis loops, Dash lines were calculated using (2). Solid lines were calculated using (5).

### 4. Model

Let us consider firstly an anisotropic superconducting plate with strong anisotropy, such that any current circulations along the *c* axis of the plate (perpendicular to the wider surface that is the *ab* plane) are negligible. The external magnetic field *H* induces currents in the *ab* plane of the plate, and the corresponding magnetization *M*$^*$ heads along the *c* axis. The related projection of *H* on the *c* axis is $H^* = H \cos \alpha$, here $\alpha$ is the angle between *H* and *c*. With the schematic presented above (Fig. 1), two projections of *M*$^*$ are measured:

$$M_x(H) = M^*(H^*) \cos \alpha,$$
$$M_y(H) = -M^*(H^*) \sin \alpha. \qquad (1)$$

The minus in the expression of $M_y$ takes into account that *M*$_y$ directs along the *y* axis for -90° < $\alpha$ < 0° and opposite for 0° < $\alpha$ < +90° (fig. 1). As it follows from (1):

$$k_{xx0}(\alpha) = |\cos \alpha|,$$
$$k_{yx0}(\alpha) = |\sin \alpha|, \qquad (2)$$
$$k_{yx}(\alpha) = |\text{tg } \alpha|.$$

The last equality means that $k_{yx}(\alpha) = 1$ and $M_{\text{irr},y} = M_{\text{irr},x}$ at $\alpha = \pm 45°$. However the data obtained from the magnetization hysteresis loops provide another value for the sample: $k_{yx}(\pm 45°) = 0.73 \pm 0.03$. Indeed, the data points for $k_{yx0}(\alpha)$ and $k_{yx}(\alpha)$ are described by equations (2) only for $|\alpha| < 30°$ (figures 3c,d). The reason of this divergence is supported to be the crystallite misalignment.

Any crystallite in the textured sample is characterized by the orientation angle θ, which is the angle between the *ab* plane of the crystallite and the texturing plane (see fig. 5), $-90° \leq \theta \leq +90°$. Let us assume that all the crystallites have the same $M^*(H^*)$ dependence, and the distribution of orientation angles is described by the normal distribution function:

$$f(\theta, \sigma) = N/\sigma \, \exp(-0.5(\theta/\sigma)^2), \qquad (3)$$

where σ is the standard deviation, $N$ is a normalizing coefficient. The orientation of any crystallite in the rotated sample is given by the angle $\alpha + \theta$. It should be noted that the orientation $\alpha + \theta$ is equivalent to $\alpha + \theta \pm 180°$. Then the magnetization projections of the textured sample are determined by

$$M_x(H) = \frac{1}{\pi} \int_{-90°}^{+90°} M^*(H^*) \cos(\alpha + \theta) f(\theta, \sigma) S(\alpha + \theta) d\theta,$$

$$M_y(H) = -\frac{1}{\pi} \int_{-90°}^{+90°} M^*(H^*) \sin(\alpha + \theta) f(\theta, \sigma) S(\alpha + \theta) d\theta, \qquad (4)$$

where $S(\varphi)$ is a function such that $S(\varphi) = 1$ for $|\varphi| \leq 90°$ and $S(\varphi) = -1$ for $|\varphi| > 90°$. This function is needed to account all upturned crystallites. The crystallites with any value of $\alpha + \theta$ have the same sign of their contribution to $M_x$. At the same time, the signs of the contribution to $M_y$ are opposite for the crystallites with $\alpha + \theta < 0$ and $\alpha + \theta > 0$ (see figure 5).

The value of $M^*$ is independent of α only at $H = 0$. It allows us to express the $k_{xx0}$, $k_{yx0}$, and $k_{yx}$:

$$k_{xx0} = \left| \int_{-90°}^{+90°} \cos(\alpha+\theta) f(\theta, \sigma) S(\alpha+\theta) d\theta \middle/ \int_{-90°}^{+90°} \cos(\theta) f(\theta, \sigma) d\theta \right|,$$

$$k_{yx0} = \left| \int_{-90°}^{+90°} \sin(\alpha+\theta) f(\theta, \sigma) S(\alpha+\theta) d\theta \middle/ \int_{-90°}^{+90°} \cos(\theta) f(\theta, \sigma) d\theta \right|, \qquad (5)$$

$$k_{yx} = \left| \int_{-90°}^{+90°} \sin(\alpha + \theta) f(\theta, \sigma) S(\alpha + \theta) d\theta \middle/ \int_{-90°}^{+90°} \cos(\alpha + \theta) f(\theta, \sigma) S(\alpha + \theta) d\theta \right|.$$

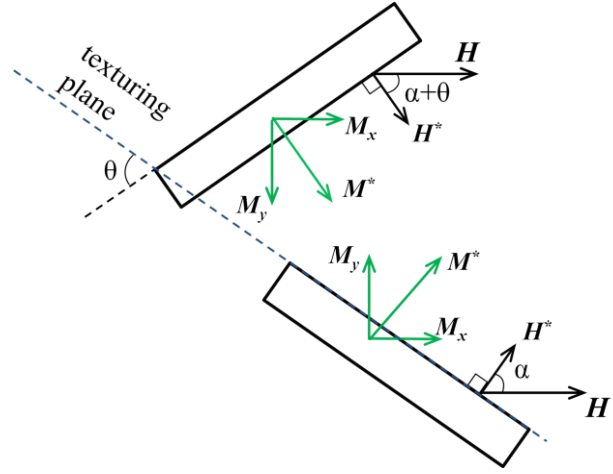

Fig. 5. Anisotropic superconducting crystallites in magnetic fields. The upper crystallite is rotated at some angle $\theta > 0$ relative to the texturing plane. The lower crystallite has $\theta = 0$.

## 5. Discussion

We used (5) to fit the normalized magnetizations $k_{xx0}(\alpha)$, $k_{yx0}(\alpha)$, and $k_{yx}(\alpha)$ on figures 4b,c,d. The single fitting parameter is used to be $\sigma$. The dash lines on figures 4b,c,d, which were calculated using (2), correspond to $\sigma = 0°$. The best agreement between the data points and computed curves (solid lines) was achieved for $\sigma = 28°$. It appears that maxima of $k_{yx0}(\alpha)$ and $k_{yx}(\alpha)$ occurs at $|\alpha| > 45°$. Also the normal distribution (3) with $\sigma = 28°$ describes successfully the bar chart obtained from SEM images (figure 2c).

Discrepancy between the data points and the curves for the well-ordered case ($\sigma = 0°$, dashed lines on figures 4b,c,d) may be used to estimate the value of $\sigma$. It is supported that the value of $k_{xx0}(90°)$ is most convenient for this estimation. It should be noted that the anisotropy coefficient $\gamma$, which used in previous works about the related textured Bi2223 [10, 12, 13], corresponds to $1/k_{xx0}(90°)$. Figure 6 shows the dependence of $k_{xx0}(90°)$ on $\sigma$ (dash line) that was computed using (5). The arrows demonstrate the determination of $\sigma$ for the considered sample ($k_{xx0}(90°) = 0.4$).

The magnetization measurements for two orientations ($\alpha = 0°$ and $\alpha = 90°$) are required to obtain the values of $k_{xx0}(90°)$, $\gamma$, and $\sigma$ [6]. The same result can be attained from $k_{yx}(\alpha)$ using only one orientation $45° \leq |\alpha| < 90°$. An example of such estimation for $k_{yx}(45°) = 0.73$ is presented on figure 6, where the dependence of $k_{yx}(45°)$ on $\sigma$ was computed using (5) (solid line). As it is seen on figure 6, the values of $\sigma$ estimated from $k_{xx0}(90°)$ and $k_{yx}(45°)$ coincide with the fitting result $\sigma = 28°$.

The another parameter, the magnetic misalignment angle $\theta^*$, was used to characterize anisotropy in previous works [13, 21]. As it was derived in [13], the value of $\theta^*$ is determined from the experimental magnetization loops as $\theta^* = \mathrm{arccot}(M_{\mathrm{irr},x}(0°)/M_{\mathrm{irr},x}(90°)) = \mathrm{arctg}(k_{xx0}(90°))$.

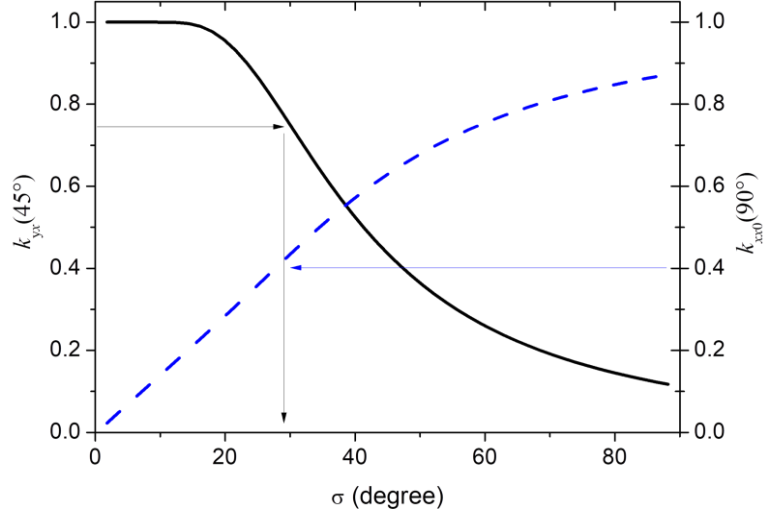

Fig. 6. The normalized irreversible magnetization for different values of the standard deviation. The values of $k_{xx0}(90°)$ (dash line) and $k_{yx}(45°)$ (solid line) are computed using (5). The arrows demonstrate the determination of σ for the investigated sample.

We suppose that $\theta^*$ corresponds to the boundaries between the quartiles of the orientation angle distribution. Providing σ = 28° in (3), one obtains $\theta^* = 21.5°$ (see left and right vertical dash lines on figure 2c). The similar value of $\theta^*$ was early estimated for the texture Bi2223 ceramics [13].

In paper [19], the Lotgering factor (LF) as an index of crystallographic orientation was correlated with σ. The value of σ = 28° corresponds LF about 0.6. For the same material the Lotgering method from X-ray diffraction data gave LF ≈ 0.98 [9]. This mismatch is explained by the dependence of texture on the distance from the sample surface. It may be suspected that the technique [9], which was used to obtain the textured samples, does not align well crystallites into the sample inner.

**Conclusion**

The anisotropic magnetization of the textured polycrystalline Bi2223 superconductor was investigated. The magnetization projections parallel and perpendicular to the magnetic field were simultaneously measured. The low value of anisotropy (γ = 2.5) was confirmed to be the result of the crystallite misalignment. It was established that the distribution of the crystallite misalignment angles can be obtained from microscopic images and from magnetization data for different orientations of the sample and the external magnetic field. Assuming the normal distribution of the crystallite misalignment angles, the standard deviation can be estimated from the magnetization data for the parallel (α = 90°) and perpendicular (α = 0°) orientations or from the $M_x$ and $M_y$ projections for the single orientation (preferably α = 45°).

**Acknowledgments** Scanning electron microscopy and magnetic measurements were carried out at Krasnoyarsk Regional Center of Research Equipment of Federal Research Center «Krasnoyarsk Science Center SB RAS».

**Funding** This work was supported by the Russian Foundation for Basic Research and the Government of the Krasnoyarsk Territory, Krasnoyarsk Territorial Foundation for Support of Scientific and R&D Activities, project "Superconducting properties of YBCO incorporated by paramagnetic rare-earth elements" No. 20–42-240008.

**Data availability** The data that support the findings of this study are available from the corresponding author upon reasonable request.

**Declarations**

**Conflict of Interest** The authors declare no competing interests.

### References

1. Kharissova, O. V., Kopnin, E.M., Maltsev, V. V., Leonyuk, N.I., León-Rossano, L.M., Pinus, I.Y., Kharisov, B.I.: Recent Advances on Bismuth-based 2223 and 2212 Superconductors: Synthesis, Chemical Properties, and Principal Applications. Crit. Rev. Solid State Mater. Sci. 39, 253–276 (2014). https://doi.org/10.1080/10408436.2013.836073
2. Hensel, B., Grasso, G., Flükiger, R.: Limits to the critical transport current in superconducting $(Bi,Pb)_2Sr_2Ca_2Cu_3O_{10}$ silver-sheathed tapes: The railway-switch model. Phys. Rev. B. 51, 15456–15473 (1995). https://doi.org/10.1103/PhysRevB.51.15456
3. Han, G., Ong, C.: Dissipation near in a textured silver-clad tape. Phys. Rev. B 56, 11299–11304 (1997). https://doi.org/10.1103/PhysRevB.56.11299
4. Pérez-Acosta, L., Govea-Alcaide, E., Noudem, J.G., Machado, I.F., Masunaga, S.H., Jardim, R.F.: Highly dense and textured superconducting $(Bi,Pb)_2Sr_2Ca_2Cu_3O_{10+\delta}$ ceramic samples processed by spark-plasma texturing. Ceram. Int. 42, 13248–13255 (2016). https://doi.org/10.1016/J.CERAMINT.2016.05.122
5. Pérez-Acosta, L., Govea-Alcaide, E., Rosales-Saiz, F., Noudem, J.G., Machado, I.F., Jardim, R.F.: Influence of the spark-plasma texturing conditions on the intragranular features of Bi-2223 ceramic samples. J. Mater. Sci. Mater. Electron. 30, 6984–6992 (2019). https://doi.org/10.1007/S10854-019-01016-6
6. García-Gordillo, A.S., Sánchez-Valdés, C.F., Sánchez Llamazares, J.L., Altshuler, E.: In-

plane anisotropy in BSCCO superconducting tapes: Transport and magnetometric criteria. Cryogenics. 109, 103102 (2020). https://doi.org/10.1016/J.CRYOGENICS.2020.103102

7. Cornejo, H.S., De Los Santos Valladares, L., Barnes, C.H.W., Moreno, N.O., Domínguez, A.B.: Texture and magnetic anisotropy of $YBa_2Cu_3O_{7-x}$ film on MgO substrate. J. Mater. Sci. Mater. Electron. 31, 21108–21117 (2020). https://doi.org/10.1007/S10854-020-04623-W

8. Pan, Y., Zhou, N., Lin, B., Wang, J., Zhu, Z., Zhou, W., Sun, Y., Shi, Z.: Anisotropic critical current density and flux pinning mechanism of $FeTe_{0.6}Se_{0.4}$ single crystals. Supercond. Sci. Technol. 35, 015002 (2021). https://doi.org/10.1088/1361-6668/AC3632

9. Petrov, M.I., Belozerova, I.L., Shaikhutdinov, K.A., Balaev, D.A., Dubrovskii, A.A., Popkov, S.I., Vasil'Ev, A.D., Mart'Yanov, O.N.: Preparation, microstructure, magnetic and transport properties of bulk textured $Bi_{1.8}Pb_{0.3}Sr_{1.9}Ca_2Cu_3O_x$ and $Bi_{1.8}Pb_{0.3}Sr_{1.9}Ca_2Cu_3O_x$+Ag ceramics. Supercond. Sci. Technol. 21, 105019 (2008). https://doi.org/10.1088/0953-2048/21/10/105019

10. Balaev, D.A., Popkov, S.I., Semenov, S.V., Bykov, A.A., Shaykhutdinov, K.A., Gokhfeld, D.M., Petrov, M.I.: Magnetoresistance hysteresis of bulk textured $Bi_{1.8}Pb_{0.3}Sr_{1.9}Ca_2Cu_3O_x$ + Ag ceramics and its anisotropy. Physica C 470, 61–67 (2010). https://doi.org/10.1016/J.PHYSC.2009.10.007

11. Gokhfeld, D.M., Balaev, D.A., Petrov, M.I., Popkov, S.I., Shaykhutdinov, K.A., Val'kov, V. V.: Magnetization asymmetry of type-II superconductors in high magnetic fields. J. Appl. Phys. 109, 033904 (2011). https://doi.org/10.1063/1.3544038

12. Gokhfel'd, D.M., Balaev, D.A., Semenov, S. V., Petrov, M.I.: Magnetoresistance anisotropy and scaling in textured high-temperature superconductor $Bi_{1.8}Pb_{0.3}Sr_{1.9}Ca_2Cu_3O_x$. Phys. Solid State. 57, 2145–2150 (2015). https://doi.org/10.1134/S1063783415110128

13. Gokhfeld, D.M., Balaev, D.A.: Magnetization Anisotropy in the Textured Bi-2223 HTS in Strong Magnetic Fields. Phys. Solid State. 62, 1145–1149 (2020). https://doi.org/10.1134/S1063783420070069

14. Daemen, L.L., Campbell, L.J., Simonov, A.Y., Kogan, V.G.: Coexistence of two flux-line species in superconducting slabs. Phys. Rev. Lett. 70, 2948 (1993). https://doi.org/10.1103/PhysRevLett.70.2948

15. Clem, J.R.: Anisotropy and two-dimensional behaviour in the high-temperature superconductors. Supercond. Sci. Technol. 11, 909–914 (1998). https://doi.org/10.1088/0953-2048/11/10/002

16. Wimbush, S.C., Long, N.J.: The interpretation of the field angle dependence of the critical

current in defect-engineered superconductors. New J. Phys. 14, 083017 (2012). https://doi.org/10.1088/1367-2630/14/8/083017

17. Mishev, V., Zehetmayer, M., Fischer, D.X., Nakajima, M., Eisaki, H., Eisterer, M.: Interaction of vortices in anisotropic superconductors with isotropic defects. Supercond. Sci. Technol. 28, 102001 (2015). https://doi.org/10.1088/0953-2048/28/10/102001

18. Obradors, X., Puig, T., Pomar, A., Sandiumenge, F., Piñol, S., Mestres, N., Castaño, O., Coll, M., Cavallaro, A., Palau, A., Gázquez, J., González, J.C., Gutiérrez, J., Romà, N., Ricart, S., Moretó, J.M., Rossell, M.D., Van Tendeloo, G.: Chemical solution deposition: a path towards low cost coated conductors. Supercond. Sci. Technol. 17, 1055 (2004). https://doi.org/10.1088/0953-2048/17/8/020

19. Furushima, R., Tanaka, S., Kato, Z., Uematsu, K.: Orientation distribution–Lotgering factor relationship in a polycrystalline material—as an example of bismuth titanate prepared by a magnetic field. J. Ceram. Soc. Japan. 118, 921–926 (2010). https://doi.org/10.2109/JCERSJ2.118.921

20. Gokhfeld, D.M., Balaev, D.A., Popkov, S.I., Shaykhutdinov, K.A., Petrov, M.I.: Magnetization loop and critical current of porous Bi-based HTS. Physica C 434, 135–137 (2006). https://doi.org/10.1016/J.PHYSC.2005.12.088

21. Lehndorff, B., Hortig, M., Piel, H.: Temperature-dependent critical current anisotropy in Bi-2223 tapes. Supercond. Sci. Technol. 11, 1261–1265 (1998). https://doi.org/10.1088/0953-2048/11/11/011